\documentclass[twocolumn,pra,aps]{revtex4}
\usepackage{graphicx}

\renewcommand{\@}[1]{\sqrt{#1}}

\def\be{\begin{eqnarray}}
\renewcommand{\le}[1]{\label{#1}\end{eqnarray}}
\def\ee{\end{eqnarray}}

\def\ffract#1#2{\raise .35 em\hbox{$\scriptstyle#1$}\kern-.25em/
\kern-.2em\lower .22 em \hbox{$\scriptstyle#2$}}

\begin{document}

\title{At World's End:\\ Where Complementarity and Irreversibility meet in the Black Hole}

\author{Giovanni  Arcioni$^1$ and Antoine Suarez$^2$ }

\address{$^1$Statistical and Mathematical Modeling, Noustat S.r.l, Via Pirelli 30, 20124 Milano, Italy\\$^2$Center for Quantum Philosophy, P.O. Box 304, CH-8044 Zurich, Switzerland \\
g.arcioni@noustat.it, suarez@leman.ch, www.quantumphil.org}

\date{January 15, 2009}

\begin{abstract}

It is argued that a slight modification of the complementarity principle may help to overcome paradoxes about the observer who falls through the event horizon.\\


\end{abstract}


\maketitle

\section{Introduction}

By assuming that quantum information falling into a black hole comes out encoded in the Hawking radiation, one avoids the information loss paradox, that is, the conclusion that the observer outside the black hole watches a non-unitary evolution of a quantum state, in sharp conflict with the rules of quantum physics. However, if a quantum state crosses the event horizon and at the same time it escapes the black hole encoded in the outgoing radiation, it means that the state has been cloned, and this is again at odds with the principles of quantum physics.

To avoid the cloning paradox one can invoke the principle of complementarity \cite{gerard1}, which basically states that cloning is possible provided it cannot be observed. If the external observer Bob can receive quantum information from the radiation but cannot get hold of the original state after its crossing the event horizon, no contradiction results. In other words duplication of information behind the horizon and in the Hawking radiation and similar contradictions never occur since, according to the complementarity principle, the black hole interior is not in the causal past of any observer who is able to measure the information content of the outgoing radiation.

Nevertheless, for a large enough  black hole one can assume, in accordance with the laws of  General Relativity, that Alice remains alive for a long time after crossing the event horizon and therefore she is able to perform quantum experiments. Suppose then that Bob recovers from the radiation a copy of the quantum information Alice carries when she crosses the event horizon. Thereafter he also crosses the horizon. Alice could thus send to Bob her quantum information. One concludes that he would be owing two copies of the same quantum state: quantum information would have been cloned.

A possible way out of this new conundrum consists in postulating a black hole's \emph{information retention time}. If it takes a certain time to Bob to collect Alice's q-bit from the radiation, then his jump into the black hole can be delayed enough so that he meets the singularity before receiving any message from Alice. Then Bob is prevented from observing quantum cloning. \cite{SuTh, HP}

A different way is the hypothesis of a ''final state''. According to it the information has first to reach the singularity before becoming ``reflected" and subsequently encoded in the outgoing radiation through a process similar to teleportation. When Bob collects Alice's q-bit from the radiation, the original state doesn't exists any more within the black hole.\cite{HoMa}

However, complementarity leads to other quandaries in case of the infalling observer.

Assuming that observers beyond the horizon are able to perform quantum experiments is only a particular case of the more general framework  that observers beyond the horizon are able to generate new information, as for instance sort of master work of literature or some relevant scientific discovery. In this context the question arises: will this new information become destroyed during the process of black hole evaporation, or will it also at the end become encoded in the radiation? According to the complementarity principle what happens beyond the horizon does not belong to the physical reality for the observer outside. As a consequence new information created by the observer inside does not become encoded in the outgoing radiation. This is baffling because the observer outside knows very well that, if he jumps into the very massive black hole, he will \emph{observe} de facto many things going on.

Moreover, if information about Alice's death due to tidal forces longtime after she crossed the horizon  goes lost forever then Hawking radiation will reveal the full quantum information defining Alice alive, i.e. the external observer compels her resuscitation.

This article proposes a new way for solving the mind boggling paradoxes of the infalling observer, and in particular quantum cloning, by using a slight modified complementarity principle. In Section II we first present the thought-experiment about quantum cloning. In Section III we define the concept of \emph{death boundary}, and discuss different outcomes according to the possible locations of the death boundary with respect  to the event horizon. In Section IV we introduce the concept of \emph{world's end} and argue that a slight modification of the complementarity principle may help to solve oddities regarding the infalling observer. In Section V we show that the new complementarity relates to irreversibility. Section VI summarizes the conclusions.

\section{Quantum cloning in a black hole}

\begin{figure}[t]
\includegraphics[width=70 mm]{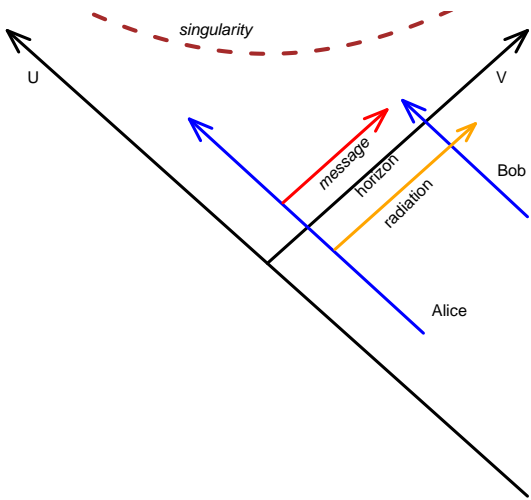}
\caption{The Hayden-Preskill thought-experiment on quantum cloning: Alice crosses the event horizon carrying a quantum-bit of information. The subsequent Hawking photons are collected by Bob, who reconstructs Alice's q-bit and thereafter jumps into the black hole too. Then Alice sends her q-bit to Bob by means of a photon message. If Bob has time to receive this message he gets to observe quantum cloning of Alice's original q-bit.}
\label{f1}
\end{figure}

Consider the situation sketched in Figure \ref{f1}. Alice falls into a Schwarzschild black hole and crosses the event horizon carrying a quantum-bit of information. The subsequent Hawking photons are collected by Bob, who is soaring outside the black hole at a certain distance from the horizon. He can thus  in principle reconstruct Alice's bit. Once Bob has collected this information he jumps into the black hole as well.
Assuming a very massive black hole, Alice and Bob can continue to perform experiments after crossing the horizon. Suppose Alice sends her q-bit to Bob by means of a photon message. By receiving it Bob gets to observe quantum cloning of Alice's original q-bit. This would contradict the principles of quantum mechanics \cite{HP, SeSu}.

To analyze the experiment \cite{HP} it is convenient to use the Kruskal coordinates $U$ and $V$. If $R$ denotes the Schwarzschild radius (i.e. $R=2MG/c^2$), then $U$, $V$ are related to the Schwarzschild coordinates $r,t$ by the following equations:

- Inside the horizon, i.e. for $0\leq r\leq R$:
\be
U&=&e^{r/2R}\left(1-\frac{r}{R}\right)^{\frac{1}{2}} e^{-t/2R}\nonumber\\ V&=&e^{r/2R}\left(1-\frac{r}{R}\right)^{\frac{1}{2}}  e^{t/2R}
\label{UV}
\ee\\

- Outside the horizon, i.e. for $R\leq r\leq \infty$:
\be
U&=&-e^{r/2R}\left(\frac{r}{R}-1\right)^{\frac{1}{2}} e^{-t/2R}\nonumber\\ V&=&e^{r/2R}\left(\frac{r}{R}-1\right)^{\frac{1}{2}}  e^{t/2R}
\label{UV2}
\ee\\

The curves for constant $r$ are given by:
\be
UV&&=e^{r/R}\left(1-\frac{r}{R}\right)
\ee

The singularity ($r=0$) is given by $UV=1$, and the event horizon ($r=R$) by $U=0$.

The proper time Alice measures between crossing the horizon ($U=0)$ at $V=V_A$ and reaching coordinate $U$ is given by

\be
	\tau_A = k R V_A U
	\label{tau}
\ee

where $k$ is a constant that depends on Alice's initial data. When Alice falls freely from infinite $k=1/e$. We assume Alice freely falls from near horizon so that in this case $k$ is O(1).

Notice that the total proper time for an observer to fall from rest $r=R$ to $r=0$ is \emph{finite} and given by:

\be
	\tau \sim R
\ee

Consider now that Bob requires a time $\Delta t$ to collect relevant information from the radiation, and then crosses the horizon at $V_B$. From (\ref{UV2}) one gets then

\be
	\frac{V_B}{V_A}=e^{\Delta t/2R}
	\label{Vb}
\ee

To observe quantum cloning Bob has to receive Alice message sending her q-bit before he hits the singularity at $UV_B=1$, i.e. at $U=\frac{1}{V_B}$. Substituting this value into (\ref{tau}) and taking account of (\ref{Vb}) one is led to:

\be
	\tau_A \sim R \frac{V_A}{V_B}=R e^{-\Delta t/2R}
	\label{tau2}
\ee

If $E$ is the energy Alice requires to sending her message to Bob, on the one hand the Uncertainty principle imposes: $\tau_A E=1$. On the other hand, the energy available to Alice is much less than the mass $M$ of the black hole. That is (in Planck units: $G=c=1$) $E<M=R/2$, and therefore $\frac{1}{\tau_A}<R/2$. Substituting into (\ref{tau2}) gives:

\be
	\frac{1}{R}e^{\Delta t/2R}<\frac{R}{2}\Rightarrow e^{\Delta t/2R}<\frac{R^2}{2}
	\label{tau3}
\ee

If one defines the Rindler retention time $\omega_{ret} = \Delta t/2R$, then (\ref{tau3}) imposes the following condition to forbid that Bob observes quantum cloning:

\be
	\omega_{ret}>lnR
	\label{omega}
\ee

Assuming that it is entanglement what causes information to become encoded in the Hawking radiation \cite{pa}, one can assume that information deposited in the black hole prior to the ``half-way'' point (before half of the entropy has been radiated away) remains concealed until the evaporation of the black hole reaches this point, it then emerges quickly for a moment, and thereafter the evaporation proceeds by emitting radiation without information. In Planck units the time necessary to radiate half the entropy of the black hole is of order $M^3$, that is, a Rindler time of order $R^2$ which exceeds the bound (\ref{omega}) by a too large amount. Complementarity would look more compelling if it just barely escapes inconsistency \cite{SeSu}.

However, if Alice jumps into the black hole when the evaporation has already proceeded beyond the ``half-way point'', and Bob has watched the system during a long time before, then Alice's quantum information may be revealed in the Hawking radiation very rapidly. Assuming that black holes scramble (thermalize) information very fast, the estimate of a black hole's information retention time is just barely compatible with the bound (\ref{omega}), what is considered a more gratifying situation \cite{HP,SeSu}.

We would like to stress that according to this explanation the information is supposed to remain ``stored'' at the ``half-way'' hyper-surface.

The ``final state'' explanation  \cite{HoMa} meets even more severe difficulties. The model is based on the assumption that reaching the singularity the information becomes somehow ``teleported'' into the outgoing radiation, so that the original state disappears before Bob can collect information from the radiation. To this aim a measurement just at the singularity is required. One has then the problem of defining how such a measurement occurs, since there are no observers around to do any measurements. Additionally any true measurement at the singularity would unavoidably lead to information loss. Thus the model works without true measurements. But then it neglects the interaction between the collapsing matter and the infalling Hawking radiation inside the event horizon \cite{GP} - a necessary condition to achieve the supposed ``teleportation''.

\begin{figure}[t]
\includegraphics[width=70 mm]{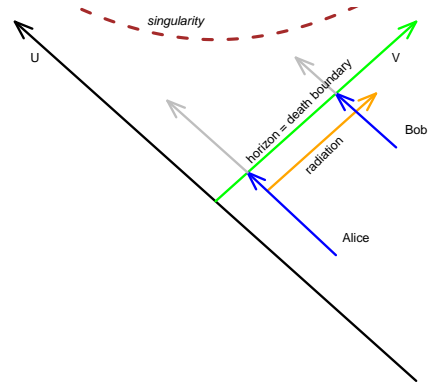}
\caption{The event horizon overlaps with the death boundary, i.e. the boundary at which human observers are killed by the tidal forces. Even if the radiation starts to carry information immediately when observers cross the horizon, these will never be able of observing cloning. In this case the event horizon marks a world's end, i.e. the end of the physical reality for any human observer.}
\label{f3}
\end{figure}

\section{The death boundary}

Consider again the infalling observer Alice: she only experiences tidal forces in her reference frame. If the black hole is not massive enough, tidal forces can become so large outside the horizon that Alice will be destroyed even before crossing it. By contrast, if the black hole is massive enough tidal forces at the horizon are negligible and Alice experiences nothing special when crossing it.

But as far as Alice keeps on falling inside the black hole tidal forces strongly increase and she is going to be destroyed \emph{before} getting to the singularity.

Tidal forces are measured by the components of the Riemann curvature tensor with
respect to Alice orthonormal reference frame (i.e. Alice's geodesic equation). One can show that if the body falling towards the singularity has mass $\mu$, height $L$ and width $w$, the radial component of the tidal force will produce \cite{Misner} a pressure with value:

\be
T_{\rho \rho} = \frac{ \mu M G L}{4 w^2 r^3}
\label{pressure}
\ee

Similar expression holds for the angular components, up to a different numerical factor.

What is important to note in our context, however, is the fact that there is a  critical value of the pressure $p_{crit}$ at which an irreparable damage occurs in any human brain. In other words, this critical value defines death, i.e. a damage that humans are not able to repair. In some sense $p_{crit}$ can be interpreted as a new constant of Nature whose real value remains inaccurately measured for the time being. We take the value of 10 atm just as a first rough estimate.

We call \emph{death boundary} the hyper-surface at which $T_{\rho \rho}$ reaches the critical value while $R^+$  is the corresponding value of the Schwarzschild coordinate $r$. From (\ref{pressure}) it follows:

\be
R^+ = \left(\frac{1}{p_{crit}} \frac{ \mu M  G L}{4 w^2}\right)^{1/3}
\label{R+}
\ee

Referring to the previous section this implies that the reasoning leading to a bound for retention time should be done not with the singularity at $r=0$, but with the curve characterizing the \emph{death boundary} at $r=R^+$, where Bob gets killed. Though this does not give a significant difference for calculating a retention time, it opens the door to another way of solving the paradoxes.

One has different possible locations for the death boundary: inside, at or outside the event horizon:

Consider first the case sketched in Figure \ref{f3}, where the death boundary overlaps with the horizon $R=R^+$. Even if the radiation starts to carry information immediately when observers cross the horizon, these will never be able of observing cloning. We assume for the human brain stem approximative values of: $p_{crit}$= 10 atm , $\mu = 20gr$, $L=5cm$, $w=2cm$. Imposing $R=R^+$ formula (\ref{R+}) gives a value $M=6.74339 \times10^{35}$ (i.e. about $339$ Solar masses) and $R^+=1000 km$.

\begin{figure}[t]
\includegraphics[width=70 mm]{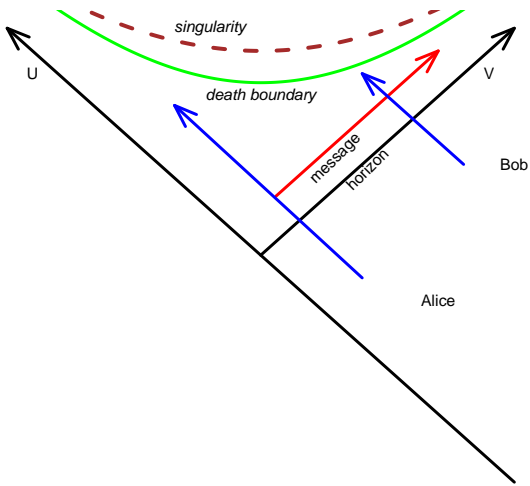}
\caption{The death boundary lies within the event horizon. If the radiation starts to carry information when the evaporation proceeds beyond the death boundary, then quantum cloning remains forbidden. The death boundary marks a world's end.}
\label{f2}
\end{figure}

Suppose now the case sketched in Figure \ref{f2}, where the death boundary lies between the singularity and the event horizon, i.e. $0<R^+<R$. When the evaporation process goes on, the event horizon will grow lower as fast as $M$. So there is a moment where one reaches again the situation $R=R^+$. If one assumes that information remains concealed till the evaporation proceeds beyond this point, then observers will never be able of observing quantum cloning either.

\begin{figure}[t]
\includegraphics[width=70 mm]{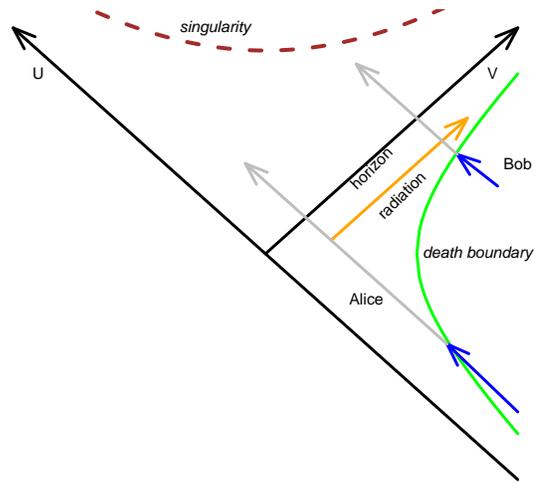}
\caption{The death boundary lies outside the event horizon. If the radiation starts to carry information when the observers cross the horizon, these will never be able of observing cloning. The event horizon marks a world's end.}
\label{f4}
\end{figure}

Consider finally the case sketched in Figure \ref{f4}, where the death boundary lies outside the event horizon, i.e. $R<R^+<\infty$. Like in the case $R=R^+$, even if the radiation starts to carry information immediately when observers cross the horizon, they will never be able of observing cloning. Notice however that in this case the radiation could not start encoding quantum information when this crosses the death boundary, since results from eventual experiments conveniently arranged between the death boundary and the event horizon could directly reach observers at distances $r>R^+$, who then would be able to observe cloning.

\section{A slight modification of the complementarity principle}

The conclusion that quantum information stored inside  a black hole obeys the rules of  unitary quantum mechanics  and thus cannot be destroyed is based on the assumption that any quantum measurement performed beyond the horizon does not exist for the external observer. Thus, according to the complementarity principle, the event horizon marks the end of the physical reality for the external observer at infinity.

As said, this sounds particularly awkward because according to general relativity the observer at infinity has to accept that he can cross the horizon without feeling any trouble, and according quantum physics he will then observe things going on inside there.

This oddity can be overcome by a slight modified complementarity principle:

We denote \emph{world's end} any boundary to a region a human observer never will get information from (unless possibly after evaporation). If the death boundary lies inside the event horizon (Figure \ref{f2}), then this boundary marks a world's end. If the death boundary matches or lies outside the event horizon (Figures \ref{f3} and \ref{f4}), then the horizon marks the world's end.

Instead of assuming that the physical reality ends at the event horizon for observers at infinity, we assume that it ends at the \emph{world's end } for any observer. Quantum states can be reduced by measurements performed by observers within the horizon, and in this case their evolution is not unitary for external observers either. By contrast, no state reduction happens beyond a world's end for any observer.

Taking account of the analysis of the preceding section, new complementarity implies that information falling into a black hole remains concealed until the evaporation of the black hole reaches a world's end, and it then immediately emerges. Radiation emitted before or after the horizon becomes a world's end does not contain any information (in agreement with the original Hawking's assumption).

\section{New complementarity and irreversibility}

The proposed new complementarity involves the notion of death boundary and thereby relates to the concept of \emph{irreversibility}. Indeed, this concept explicitly appears in the clinical definition of death, which basically states: death occurs when the neural functions responsible for certain spontaneous movements \emph{irreversibly} break down.

In establishing death this way, we are assuming as obvious that our capacity of restoring neuronal dynamics (our repairing capability) is limited in principle, even if we don't yet know where this limitation comes from.

The same concept of irreversibly can be applied to measurement \cite{su08}. The amplification in a detector (e.g. a photomultiplier) becomes irreversible in principle at a certain level, if beyond this level an operation exceeding the human capabilities would be required to restore the photon's quantum state. When such a level is reached the detector clicks. Such a view combines the subjective and the objective interpretation of measurement: on the one hand no human observer (conscious or not) has to be actually present in order that a registration takes place; on the other hand one defines the 'collapse' or 'reduction' with relation to the capabilities of the human observer.

In our model the information characterizing living human brains goes irreversibly lost at a world's end in the same way in which information goes irreversibly lost in the detection process. Any well functioning human \emph{brain stem} breaks down at a world's end. Since the stem is the most primitive part of the brain in evolution, presumably any brain breaks down at a world's end.

In this sense a world's end can be considered a detector reducing quantum neural states.

But it can also be viewed as an ``information mirror'' \cite{HP} reflecting other types of states. Although as a mirror working in a peculiar way, since it also conceals information for a long time. Suppose a stone reaching the world's end of a huge black hole. The information defining the stone will remain concealed until the evaporation brings the horizon to match with the world's end, at which moment it gets encoded in the radiation. At the end of the day it looks as if the stone would disappear and reappear after a long delay. Actually any model assuming discrete time displays a similar behavior though for much shorter time intervals. In this sense the black hole acts like a magnifier of the fundamental unit of time.
\\
\\

\section{Conclusions and open problems}

In summary we have shown:

The world's end hypothesis goes around quantum cloning in a natural way.

Since beyond the world's end it does not even make sense to talk about physical reality for any observer, the very concept of singularity disappears.

The region between the horizon and the world's end belongs to the physical reality, and any new information arising beyond the horizon becomes encoded in the outgoing radiation, and does not go lost for the external observer.

Since information characterizing living human brains goes irreversibly lost, evaporation does not produce resuscitation.

New complementarity depends on the concept of \emph{irreversibility}. Although it seems that this concept refers to a limitation of the human capabilities, for the time being we do not know when precisely irreversibility happens, neither in death nor in measurement.

New complementarity does not address the problem of how the information remains concealed within the black-hole and becomes encoded in the outgoing radiation \cite{HP,SeSu}. Nevertheless it suggests that this problem may be related to the question of how information lasts in models working with discrete spacetime.

Reducing the problem of the infalling observer to other known unsolved problems is somewhat already a progress.


\end{document}